\documentclass[twocolumn,prb,tighten,floatfix,showpacs]{revtex4}
\usepackage{graphicx}

\def\<{\langle}
\def\>{\rangle}

\def\be{\begin{equation}}
\def\ee{\end{equation}}

\begin{document}
\preprint{cond-mat} \title{ Topological entropy of realistic quantum Hall wave functions}

\author{B. A. Friedman and G. C. Levine}

\address{Department of Physics, Sam Houston State University, Huntsville TX 77341}

\address{Department of Physics and Astronomy, Hofstra University,
Hempstead, NY 11549}

\date{\today}

\begin{abstract}The entanglement entropy of the incompressible states of a realistic quantum Hall system are studied by direct diagonalization. The subdominant term to the area law, the topological entanglement entropy, which is believed to carry information about topologic order in the ground state, was extracted for filling factors $\nu = 1/3$, $\nu = 1/5$ and $\nu = 5/2$.  The results for $\nu = 1/3$ and $\nu = 1/5$ are consistent with the topological entanglement entropy for the Laughlin wave function.  The $\nu = 5/2$ state exhibits a topological entanglement  entropy consistent with the Moore-Read wave function.
\end{abstract}

\maketitle

\section{Introduction}
This paper is a numerical study, using direct diagonalization, of the entanglement entropy of incompressible states of quantum Hall systems.  In particular, the entanglement entropy is calculated for filling factors $1/3$ and $1/5$   in the $n=0$ Landau level and the $5/2$ state in the $n=1$ Landau level.  The primary motivation for this work is to better understand the nature of the $5/2$ state observed in experiment \cite{willett}.  Originally, the $5/2$ state was not believed to be spin polarized \cite{eisenstein}, however, with theoretical input \cite{mooreread,greiter,morf} and further experimental investigation \cite{xia} an incompressible spin polarized state was revealed.  An elegant theoretical possibility for this state is the wave function suggested by Moore and Read \cite{mooreread}; the novel feature of the Moore-Read state being the presence of non abelian fractional statistics.  As well as being of interest in its own right, states with non abelian statistics give rise to possible robust implementation of quantum computation \cite{phystoday,kitaev}.  However, the Moore-Read state is the ground state of a not very realistic Hamiltonian with a three body interaction term \cite{greiter}.  

The question remains, does the Moore-Read state contain the physics of the   $\nu = 5/2$ quantum Hall system; that is, is it in some sense ÒcloseÓ to the ground state of a model with a realistic Hamiltonian, i.e.  electrons in a magnetic  field interacting via long range Coulomb interaction?  A valuable way to address this issue is to compare the ground state wave function obtained from direct diagonalization of a realistic Hamiltonian to the Moore-Read wave function.  The results of such studies are unfortunately somewhat ambiguous \cite{rezayi,tokejain}.  In addition, even if there is a large overlap between the numerical wave function and the Moore-Read state, how does one know whether this overlap truly indicates that the long distance, low energy behavior is the same, in particular whether the numerical wave function has non abelian statistics.  Attempts to directly detect non abelian statistics in numerical systems with strictly Coulomb interactions have thus far also proven to be elusive \cite{tokejain2,wan}.  Note in reference [12], non abelian statistics were clearly observed, however, the Hamiltonian contained a mixture of Coulomb and 3-body terms. 

Recently, a novel approach, using concepts from quantum information theory, has been proposed to characterize incompressible quantum states.  Kitaev and Preskill \cite{kitaevpreskill} and Levin and Wen \cite{levinwen} have shown the sub leading  contribution to the entanglement entropy of a subsystem, the topological entanglement entropy is universal and reflects the statistics of quasiparticles of the incompressible Hall state in question.  To be more explicit, consider a two dimensional quantum many body system and spatially divide the system into two parts, the part of interest being the subsystem and the rest which is referred to as the environment.  The result of ref. [13,14] is that the entanglement entropy scales as
  \begin{equation}
S \simeq \alpha L - \gamma + O(\frac{1}{L}) + \ldots 
\end{equation}
 for topologically ordered states.  Here L is the linear size of the boundary of the subsystem and $\gamma$ is the topological entanglement entropy.  S, the entanglement entropy, more precisely defined in section II, intuitively is a measure of the quantum entanglement of the subsystem and the environment. The sub leading term, the topological entanglement entropy was shown 
 \cite{kitaevpreskill,levinwen} to be equal to the log of the total quantum dimension of the state in question.  Intuitively, \cite{fendley} $\gamma$ reflects the number of distinct quasi particle types and how the number of linear independent states, for each type, grows with the number of quasi particles in the states. 
 
It is important to note that the above expression holds asymptotically for large L, that is, a large subsystem and a very large environment.  By considering several different subsystems, it was shown that the leading contribution, which scales as the linear size of the subsystem (and is non universal) could be cancelled out and the sub leading term could be extracted.  Numerically, however, it is not easy to implement this method \cite{furukawa}.
 
          In references [17] and [18], a more practical method for quantum Hall systems, based on looking at several system sizes, was used to successfully compute the topological entanglement entropy of the Laughlin wave function for  $\nu = 1/3$   and the Moore-Read state for  $\nu = 5/2$.  In this paper we apply the method of reference [17] to the exact ground state wave functions obtained from realistic Hamiltonians by direct diagonalization.  If the value of the topological entanglement entropy extracted from a direct diagonalization calculation agrees with the value calculated for the Moore-Read state this provides  evidence that the Moore-Read state correctly describes the physics of filling factor   $\nu = 5/2$.  It is important to realize,  without other physical constraints, equality of the topological entanglement entropy is a necessary but not sufficient condition for states to be topologically equivalent.  Simply stated, if two states don't have the same value of $\gamma$, they are not equivalent.  However, equality of $\gamma$ does not necessarily imply two states are equivalent.  Therefore to completely characterize an incompressible state additional information is needed.
               
The paper is organized in the following way: In the next section, the numerical method is described, highlighting the differences from reference [17], and in the following section the numerical results are presented.  The final section is a summary and gives our conclusions.

\section{Numerical Method}

To do our direct diagonalization (DD) calculations, we work with finite square clusters with periodic boundary conditions, the flat torus geometry.  This is in contrast to reference [17] which works in the spherical geometry.  One reason we favor the torus geometry is that most published density matrix renormalization group (dmrg) calculations of quantum Hall systems, which can handle larger system sizes, are performed in this geometry \cite{shibata}(see however, Feguin et al.\cite{feguin} for dmrg in the spherical geometry)  .  Although the present work is strictly DD, we hope to lay the ground work for a future dmrg study.

To take the magnetic field into account, the Landau gauge is chosen, where the momentum in the $y$ direction is a conserved quantity and the single particle orbitals, strips of width the magnetic length, are oriented parallel to the $y$-axis.  The single particle orbitals are labeled by the $x$ guiding center coordinate or equivalently the momentum in the y direction.  Although momentum in the $x$ direction is also conserved, for simplicity, and since this symmetry has not been implemented in dmrg, we do not make use of this quantum number \cite{chakraborty1}.  The Lanczos algorithm is used to calculate the ground state in each sector of total y momentum and then the lowest energy state is selected.  For the  $\nu =1/3$ and $\nu=1/5$ fillings the state space is restricted to the $n=0$ Landau level while for the  $\nu=5/2$ filling the state space is restricted to the $n=1$ Landau level.  We again emphasize that the electrons interact via the long range Coulomb interaction periodically continued in the usual way \cite{chakraborty2}.  Due to cpu limitations the calculations were limited to system sizes smaller than or equal to 12 electrons in 36 orbitals ($\nu=1/3$), 8 electrons in 40 orbitals ($\nu=1/5$), and 16 electrons in 32 orbitals, ($\nu=5/2$).  ( the diagonalizations involve state spaces of sizes   at most $35 \times 10^6$.  It is possible, with some difficulty to extend the calculations by 1 electron for $\nu=1/5$ ).  

We now turn to the method used to compute the topological entanglement entropy.  After computing the ground state (or states, the issue of ground state degeneracy will be addressed later) the entanglement entropy is then calculated.  As in the spherical geometry, in the torus geometry, there is a natural (numerically easy) choice for the subsystem to calculate the entanglement entropy.  The subsystem chosen consists of $l$ adjacent (in $x$) orbitals, for example, for $l=2$ one can take orbitals 1 and 2 to get the 2 orbital entanglement entropy.  In a system with $N$ total orbitals (i.e. for $11$ electrons in $33$ orbitals, $N=33$) the many electron wave function has the form of a collection of coefficients $\Psi_{i_1 i_2 i_3 \dots i_N}$ where $i_1,i_2,i_3 \dots i_N$ take the values 0 or 1 and $\Psi$ is the amplitude for the state with occupancies $i_1,i_2,i_3 \dots i_N$. One then computes the $l$ orbital density matrix (an object very familiar from the density matrix renormalization group \cite{white} ). Explicitly for $l=2$,
\begin{equation}
M_{i_1 i_2 i_1^\prime i_2^\prime} = \sum_{i_3 i_4 \ldots i_N}{\Psi_{i_1 i_2 i_3 \dots i_N}  \Psi_{i_1^\prime i_2^\prime i_3 \dots i_N}}
\end{equation}
The density matrix is then diagonalized, yielding the eigenvalue $\lambda_j$ from which the entanglement entropy
\begin{equation}
-\sum_j{\lambda_j \ln{\lambda_j}}
\end{equation}
is obtained.

Since the torus geometry is used in calculating the, say, 2 particle entropy, it does not matter if one takes, the orbitals 1,2 or 30, 31 etc. However, there is a subtlety in that we are not explicitly taking conservation of x-momentum into account and due to ground state degeneracy the numerical wave function may not be translationally invariant.  To handle this problem, we have used the wave functions that are more translationally invariant in the following sense:  (take for concreteness 7 electrons in 21 states, $\nu=1/3$ ).  Calculate the "left" and "right" $l$ body entanglement entropies, taking the "left" subsystem to be orbitals $1,2,3, \ldots, l$ and the right subsystem to be $22-l, \ldots, 20,21$.  We choose the ground state where the entanglement entropies for the "left" and "right" subsystems are equal.  For example, for 7 electrons in 21 orbitals, the states with $k_y= 7,14,21$ (in appropriate units) are degenerate but only $k_y = 14$ satisfies the above criteria.  For $\nu=1/3$ and $\nu=1/5$, this criteria uniquely picks the ground state.  For $\nu=5/2$, there are certain filling factors (i.e. 14/28) where 2 ground states satisfy the equality of the "left" and "right" entanglement entropies.  In these cases, we pick the ground state with the lowest momentum. 
\begin{figure}[ht]
\includegraphics[width=6.5cm]{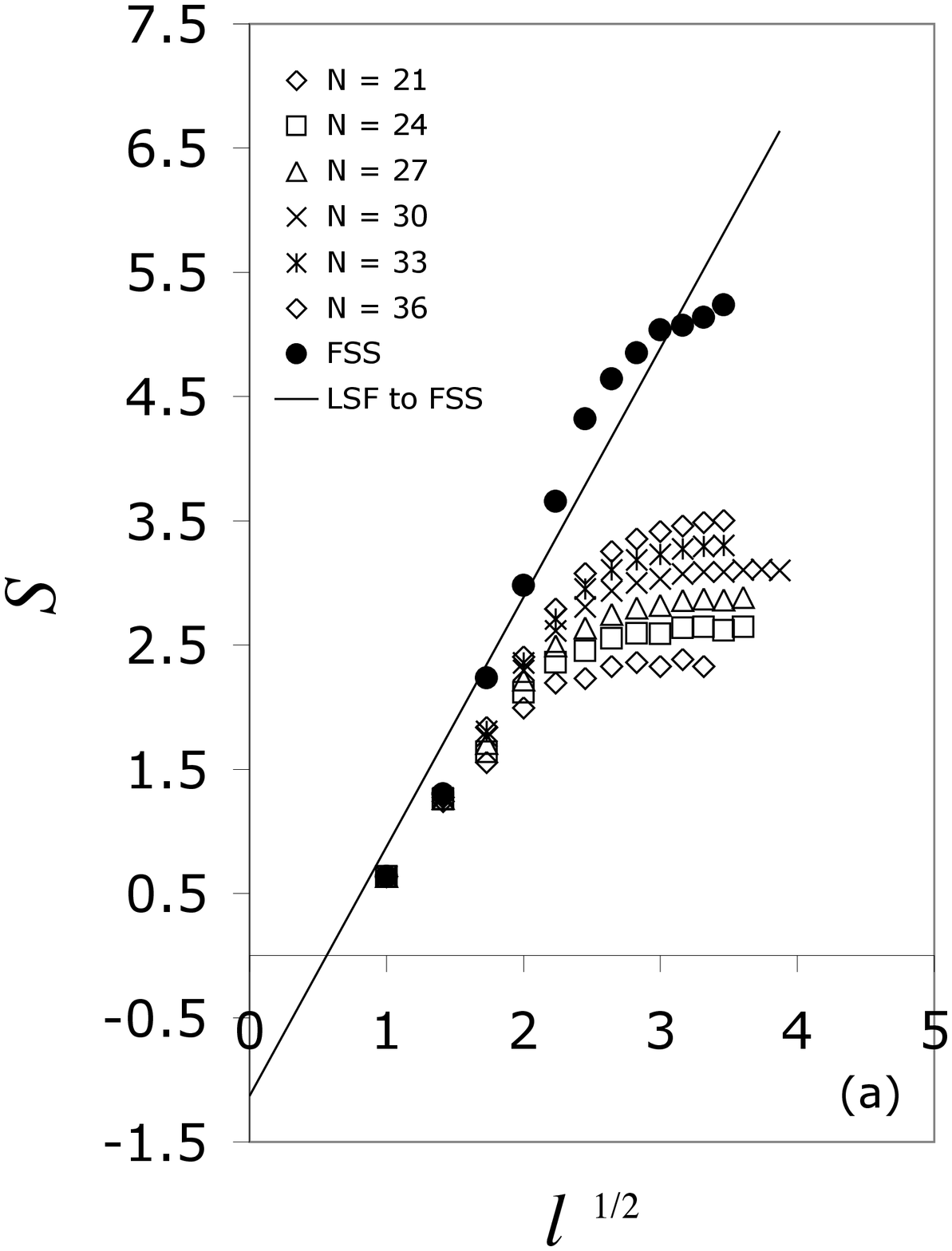}
\includegraphics[width=6.5cm]{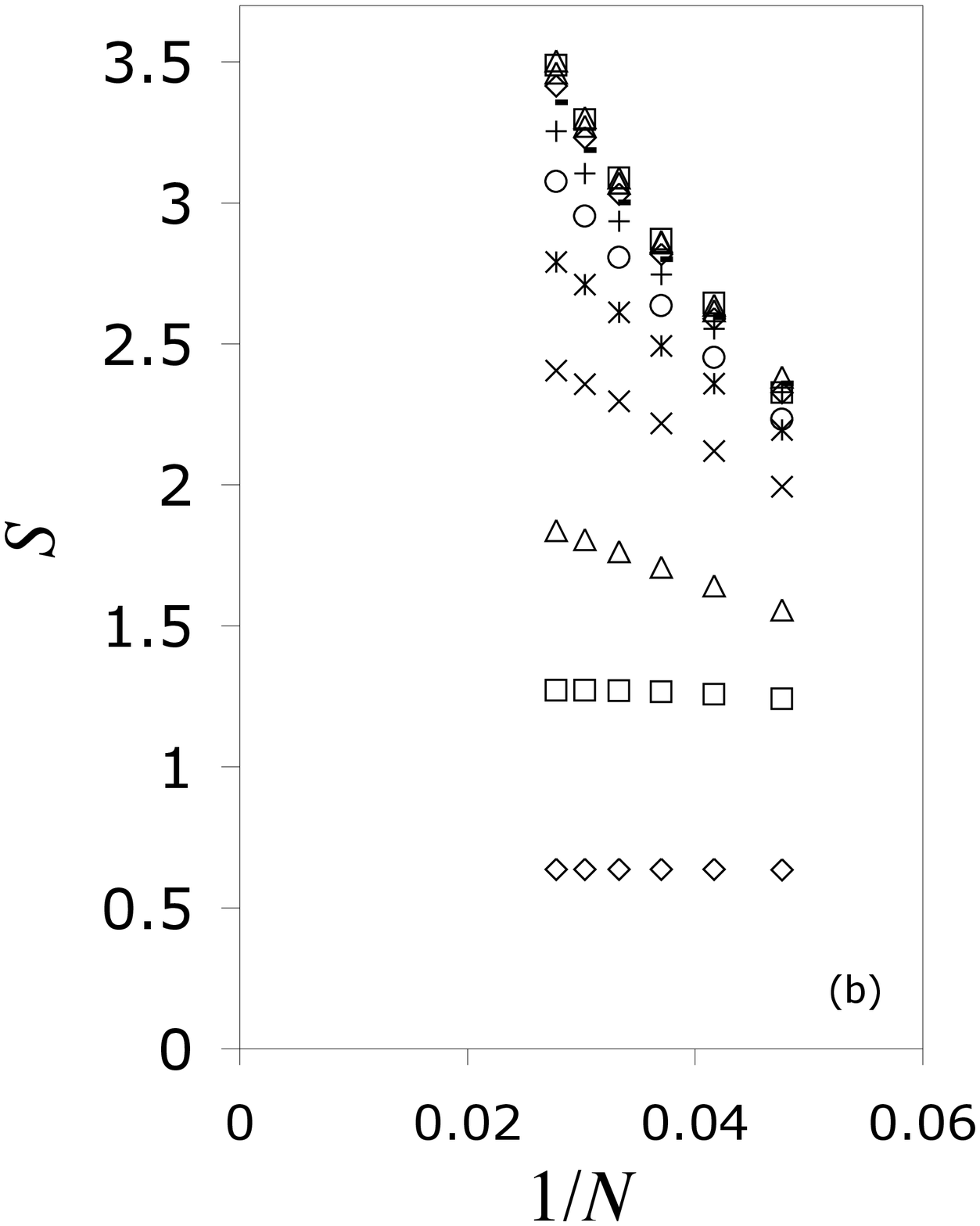}
\caption{\label{fig1} Entanglement entropy of $\nu = 1/3$ state. (a) Entanglement entropy versus $\sqrt{l}$ (where $l$ is number of orbitals comprising the sub system) for different total number of orbitals, $N=21-36$.  Finite size scaling (FSS), $N \rightarrow \infty$, results for $S$ versus $\sqrt{l}$ ($\bullet$) is shown with linear least squares fit (LSF), yielding an intercept of  $-\gamma = -1.13 \pm 0.38$. (b) Finite size scaling of entanglement entropy.  $N \rightarrow \infty$ results, shown in figure 1a, were obtained by linear LSF to $N=24,27,30,33,36$ data. Estimated uncertainty ($\sigma_S$) in $N \rightarrow \infty$ values of $S$ was $\sigma_S <  0.1$, smaller than the size of the plotting symbol ($\bullet$).}
\end{figure}

Following reference [17], from the entanglement entropies for different numbers of orbitals and different system sizes we have attempted to calculate the topological entanglement entropy. The idea is based on the asymptotic formula eq.(1).   The first term in eq. (1) is referred to as the Òarea lawÓ, entropy being proportional to the bounding area, which is one-dimensional in this case. The system and subsystem must be sufficiently large to realize the bulk behavior of the strongly correlated state, and the subsystem must be much smaller than the system to realize the area law behavior of the entanglement entropy. Consider first of all a very large system (or at least the largest system we can compute with DD) and a sufficiently large but not too large subsystem. As in ref. [17], we identify the number of Landau orbitals with the area enclosed. 
\begin{figure}[here]
\includegraphics[width=6.5cm]{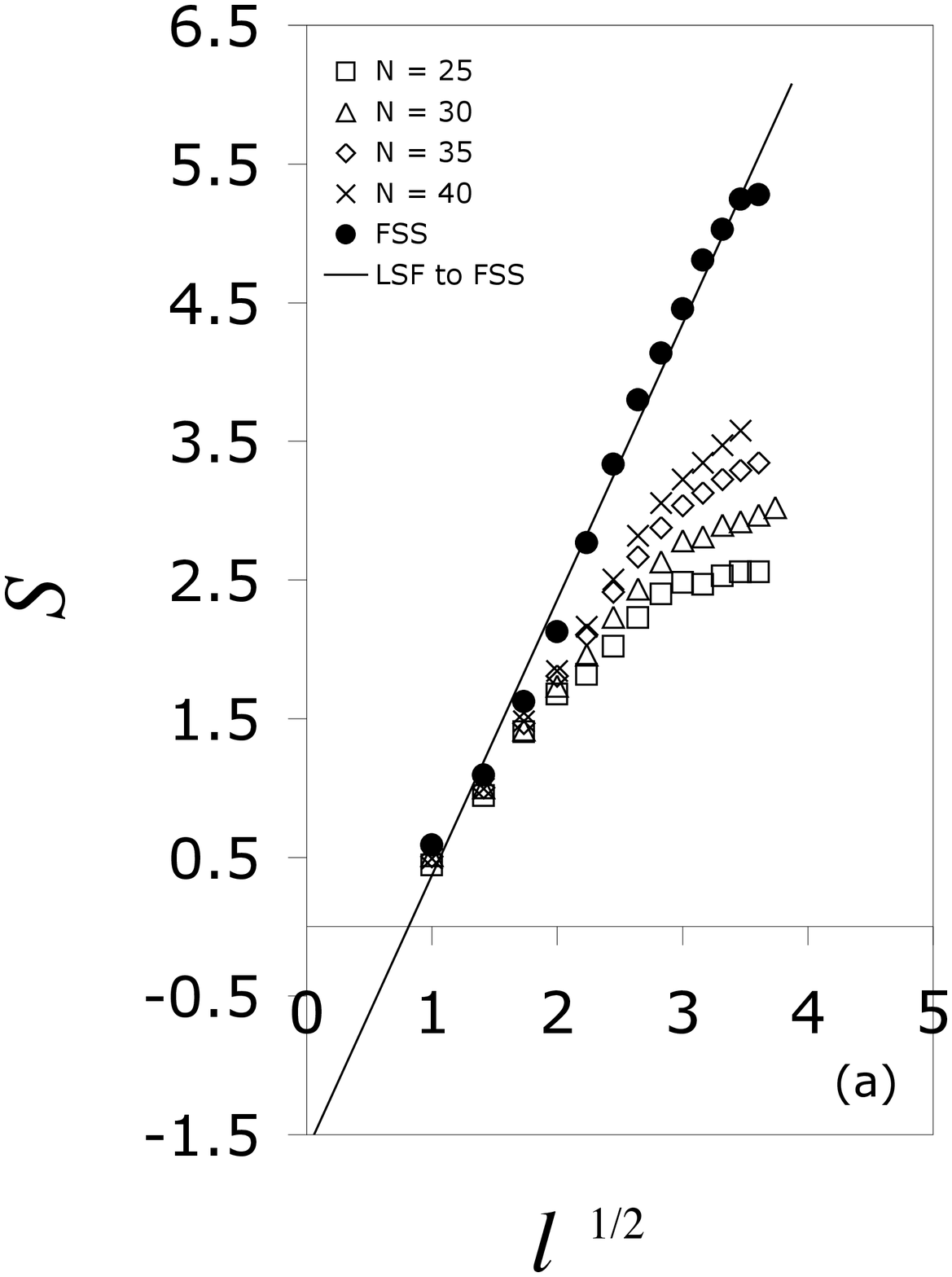}
\includegraphics[width=6.5cm]{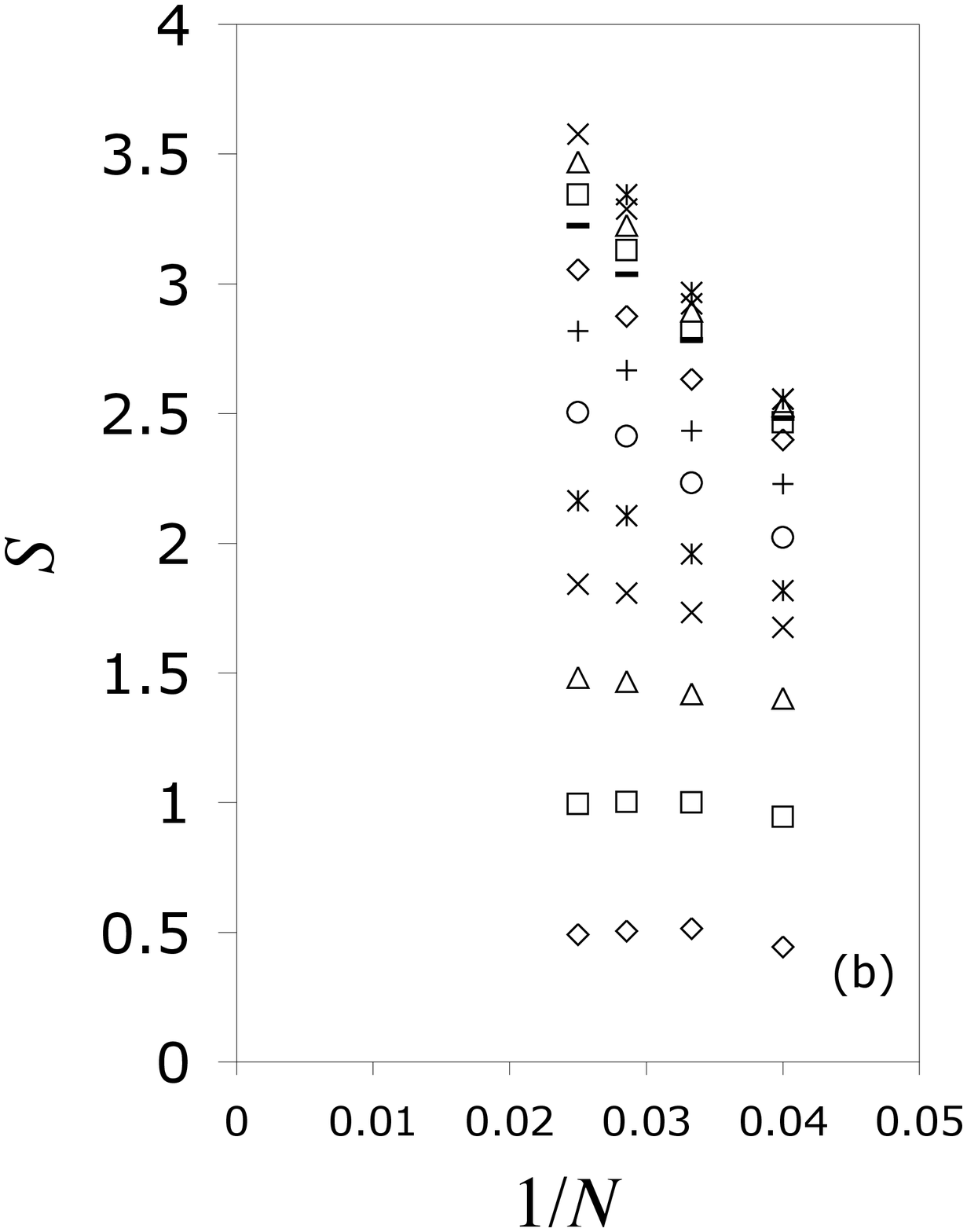}
\caption{\label{fig2} Entanglement entropy of $\nu = 1/5$ state. Same as figure 1 except: (a) $N \rightarrow \infty$ results for $S$ versus $\sqrt{l}$ ($\bullet$) yield an intercept of  $-\gamma = -1.62 \pm 0.16$  (b) $N \rightarrow \infty$ results were obtained by linear LSF to $N = 25,30,35,40$.  Estimated uncertainty in $N \rightarrow \infty$ values of $S$ was $\sigma_S <  0.15$, approximately the size of the plotting symbol ($\bullet$).  }
\end{figure}

By plotting S vs the square root of the number of orbitals, one hopes to get a straight line (reflecting the Òarea lawÓ eq. (1)); then the $y$-intercept should give minus one times the topological entanglement entropy. It is however, not easy to treat a very large system with direct diagonalization. A possibility to overcome this difficulty, is to use information from a number of system sizes. Again following [17] one can plot the $l$ orbital entanglement entropy vs $1/N$ ($N=$ number of orbitals) and try to extrapolate to an infinite system.  Our experience  indicates that a linear extrapolation in $1/N$ works better than adding nonlinear terms, so we use a simple linear extrapolation only.   The extrapolated $l$-orbital entanglement entropies are then plotted vs the square root of the number of orbitals in the subsystem ($\sqrt{l}$). If a straight line results, then the topological entanglement entropy can be extracted as the y-intercept.

A possible alternative approach to the above method is to numerically implement the method of ref. [13]  or ref. [14].  Such an approach was taken in ref. [16] and applied to the quantum dimer model.  Since the dimer model is defined on a lattice, it is easier to vary the choice of the subsystem.  However, to obtain accurate results, large system sizes had to be used (this is practical since no diagonalization is needed at the Rokshar-Kivelson point of the phase diagram) or alternatively special properties of the dimer model had to be utilized.
\section{numerical results}

Let us now examine the results of our calculation. First consider $\nu=1/3$ in figures 1a, 1b. Figure 1a is a graph of $S$, the entanglement entropy, vs the square root of the number of orbitals in the subsystem, for various system sizes. We have also shown in this figure an extrapolation to large N given by the solid circles. (See below for discussion).  For the largest system $N=36$, one sees reasonably linear behavior for up to 6 orbitals in the subsystem and a $y$-intercept of  $\approx-1$ and a topological entanglement entropy of  $\approx1$. In the spherical geometry, for  $\nu=1/3$, it is known that the topological entanglement entropy takes a value of $\gamma \simeq 0.55$. However, for the torus, the subsystem has two boundaries rather than one and even the leading term in the entanglement entropy only depends on orbitals rather close ($\sqrt{l}$) to the respective boundaries.  Hence one expects the contribution from each boundary to merely add \cite{levinwen} and give twice the value of topological entanglement entropy for the sphere.  Our numerical value $\gamma \approx1$, given the uncertainties, see figure 1a, is consistent with this expectation.

To try to get a more precise estimate in figure 1b we plot $S$, the $l$-orbital entanglement entropy, vs $1/N$  for various $l$ values.  By doing a linear least squares fit in $1/N$ for N=24,27,30,33,36 and taking the value of the line at $1/N=0$ the $l$-orbital entanglement entropy was extrapolated to large system sizes.  These values are shown by the solid circles in figure 1a.  The solid circles are then fit to a straight line and the intercept gives a topological entanglement entropy   $\gamma = 1.13 \pm 0.38$.  This is consistent with the value expected from the Laughlin state for $\nu=1/3$.  The large uncertainty is the result of the poor linearity of a curve passing through the solid circles.

In figure 2a, b  we plot analogous graphs for $\nu=1/5$.  Looking at figure 2a, the entanglement entropy, vs the square root of the number of orbitals in the subsystem, for the largest system size ($N=40$, but only $8$ electrons) we again see linear behavior for up to 6-7 orbitals.  A fit to the linear part gives a $y$-intercept of about -1 or a slightly smaller value -1.2  if we exclude $l=1$.  This is compared to an expected value of $-2\ln{\sqrt{5}} \simeq  -1.61$.  The solid circles in figure 2a are values of the entanglement entropy extrapolated from figure 2b.  A linear fit to the extrapolated values yields an intercept of   $-1.62 \pm 0.16$.  This is in excellent agreement with the value from the $\nu=1/5$ Laughlin state.  The much smaller, though still substantial, uncertainty reflects the better linearity of a curve passing through the solid circles  in comparison to  the case $\nu=1/3$.     
\begin{figure}[here]
\includegraphics[width=6.5cm]{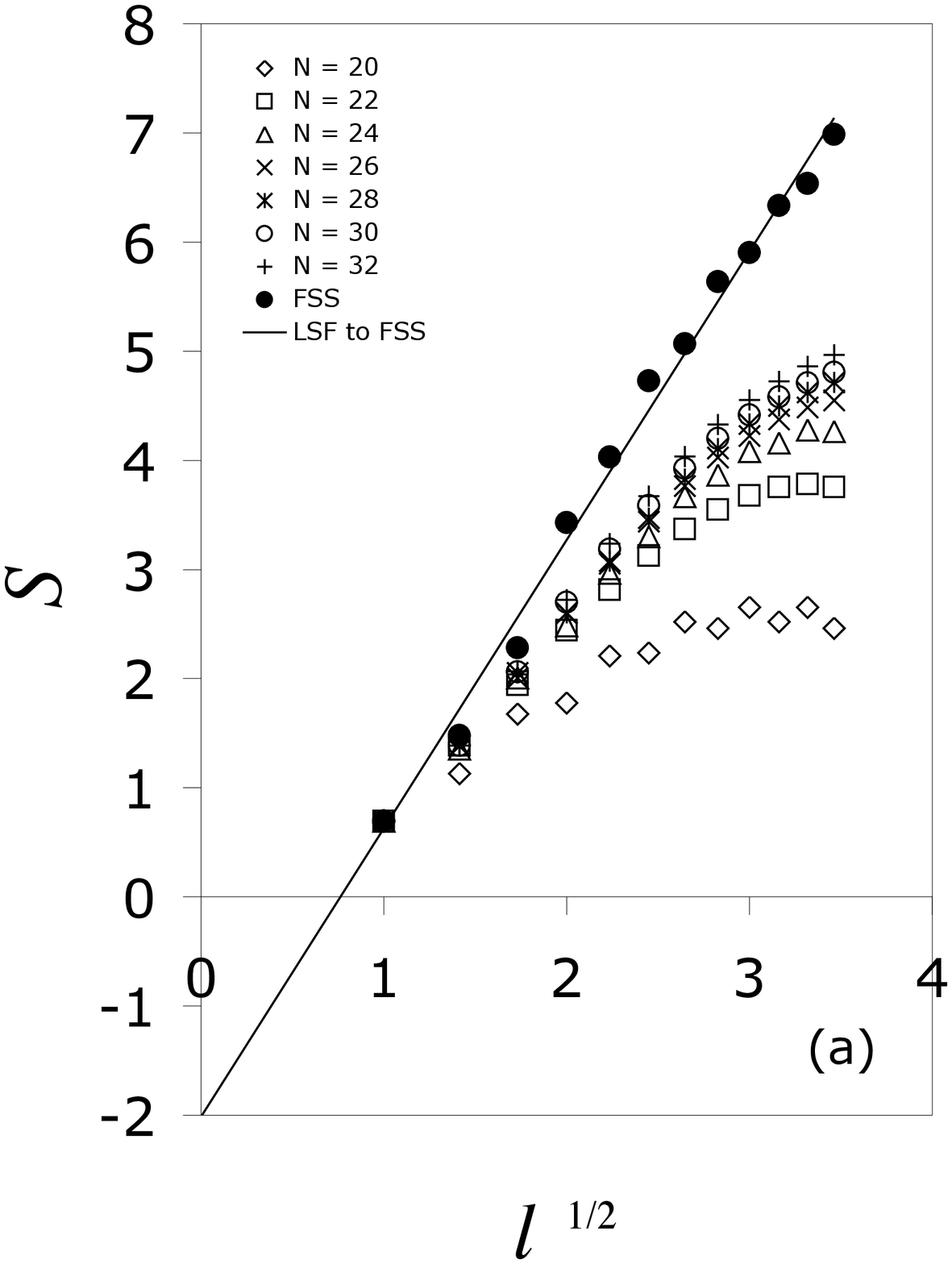}
\includegraphics[width=6.5cm]{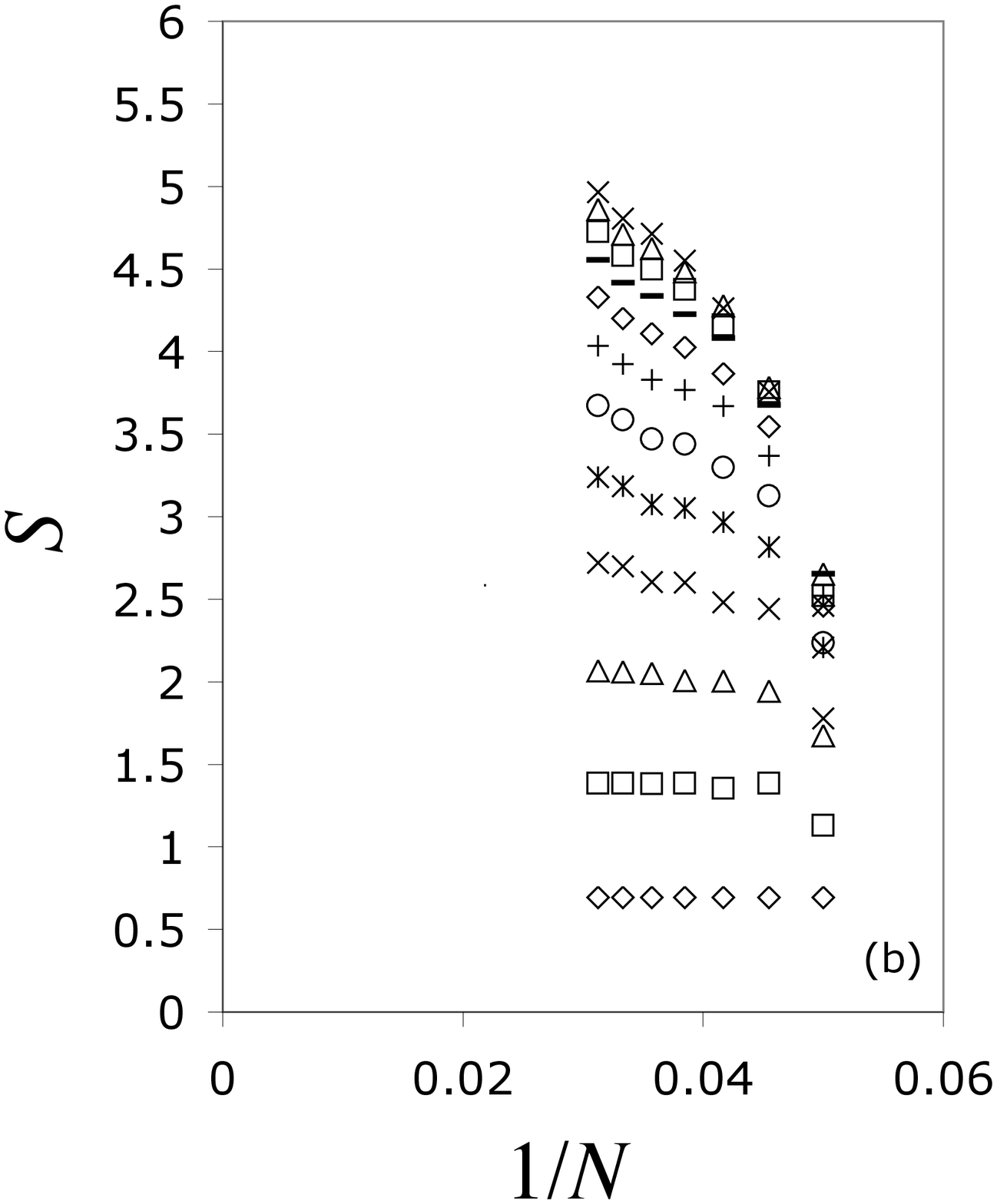}
\caption{\label{fig3} Entanglement entropy of $\nu = 5/2$ state. Same as figure 1 except: (a) $N \rightarrow \infty$ results for $S$ versus $\sqrt{l}$ ($\bullet$) yield an intercept of  $-\gamma = -2.01 \pm 0.19$  (b) $N \rightarrow \infty$ results were obtained by linear LSF to $N = 24,26,28,30,32$. Estimated uncertainty in $N \rightarrow \infty$ values of $S$ was $\sigma_S <  0.18$, approximately the size of the plotting symbol ($\bullet$).}
\end{figure}

Finally, let us investigate the $\nu=5/2$ system \cite{zozulya}.    In figure 3a we again plot , the entanglement entropy, vs the square root of the number of orbitals in the subsystem.  There is reasonably linear behavior  up to about 6-7 orbitals yielding a $y$-intercept of roughly -1.5.  On the other hand, for the Moore-Read state we expect a topological entanglement entropy of about $2 \ln{\sqrt{8}} \simeq 2.08$.   To try  to get a more precise estimate in figure 3b we plot the $l$-orbital entanglement entropy vs $1/N$ for various values of $l$.  Again a simple linear extrapolation of the five largest systems sizes is  adequate. The solid circles in figure 3a shows the extrapolated entanglement entropies.   Using a linear  fit, the extrapolated values then give a $y$-intercept of $-2.01 \pm 0.19$.  Thus, the numerical state has a topological entropy close to that of the Moore-Read state.  A possible alternative state, having the same topological entanglement entropy, is the Halperin 3-3-1 state (We thank one of the referees for pointing this out to us).  However, since we are only working with spin polarized electrons in a single layer, this state is excluded on general grounds of symmetry \cite{yoshioka} .  
  
\section{Conclusion}
In this paper, direct diagonalization, by necessity on small system sizes, has been used to calculate the entanglement entropy for the $\nu=1/3$, $\nu=1/5$ and $\nu=5/2$ quantum Hall states.  We emphasize that a realistic Hamiltonian,  long range Coulomb interaction, has been used.  At all filling fractions considered, the area law has been verified by examining the largest system size.  To accurately extrapolate the topological entanglement entropy it was necessary to extrapolate the $l$ orbital entanglement entropy to large system sizes.  For   $\nu=1/3$ and  $\nu=1/5$, this extrapolation gave results consistent with values of the topological entanglement entropy for the Laughlin state.   It should be noted that our results are consistent with an entropy corresponding to a topological ground state with {\sl two} boundaries (following from our computation in the Landau gauge) rather than a single boundary in the spherical geometry. For  $\nu=5/2$ the value of the topological term obtained was consistent with the topological term for the Moore-Read state.  We view this as a confirmation, that the incompressible state at  $\nu=5/2$, is in fact, the Moore-Read state.

We thank the referees for a number of helpful suggestions and M. P. A. Fisher and C. Nayak for helpful correspondence.  The work was supported, in part, by an award from Research Corporation CC6535 (G.L.), the Texas Advanced Research Program grant no.  003606-00050-2006 and the NSF DMR-0705048 (B.F.).


\begin{references}

\bibitem{willett} R. L. Willett et al. Phys. Rev. Lett. 59, 1776 (1987).

\bibitem{eisenstein}J. P. Eisenstein et al., Phys. Rev. Lett. 61, 997 (1988).

\bibitem{mooreread} G. Moore and N. Read, Nucl. Phys. B 360, 362 (1991).

\bibitem{greiter}M. Greiter, X.-G. Wen and F. Wilczek, Phys. Rev. Lett. 66, 3205 (1991).

\bibitem{morf} R. H. Morf, Phys. Rev. Lett. 80, 1505 (1998).

\bibitem{xia} J. S. Xia et al. Phys. Rev. Lett. 93, 176809 (2004).

\bibitem{phystoday} Physics Today, Oct 2005 p21.

\bibitem{kitaev} A. Kitaev, A. Shen and M. Vyalyi, ÒClassical and Quantum ComputingÓ, Amer. Math Soc. (2002).

\bibitem{rezayi} E. H. Rezayi, and F. D. M. Haldane Phys. Rev. Lett. 84, 4685 (2000).

\bibitem{tokejain} C. Toke and J. K. Jain, Phys. Rev. Lett. 96, 246805 (2006).

\bibitem{tokejain2} C. Toke, N. Regnault and J. K. Jain, arXiv: cond.-mat./0707.0586

\bibitem{wan} X. Wan, K. Yang and E. H. Rezayi, Phys. Rev. Lett. 97, 256804 (2006). 

\bibitem{kitaevpreskill} A. Kitaev and J. Preskill, Phys. Rev. Lett. 96, 110404 (2006).

\bibitem{levinwen}  M. Levin and X.-G. Wen, Phys. Rev. Lett. 96, 110405 (2006).

\bibitem{fendley} P. Fendley, M. P. A. Fisher, and C. Nayak, J. of Stat. Phys. 126, 1111 (2007).

\bibitem{furukawa} S. Furukawa and G. Misguich, Phys. Rev. B 75, 2114407 (2007).

\bibitem{haque} M. Haque, O. Zozulya and K. Schoutens, Phys. Rev. Lett. 98, 060401 (2007).

\bibitem{zozulya} O. S. Zozulya, M. Haque, K. Shoutens, and E. H. Rezayi, ArXiv: Cond. Mat./0705.4176

\bibitem{shibata} N. Shibata and D. Yoshioka, Phys. Rev. Lett. 86, 5755 (2001); N. Shibata, J. Phys. A. Math. Gen. 36 R 381 (2003).

\bibitem{feguin} A. E. Feguin, E. Rezayi, C. Nayak and S. Das Sarma, arXiv:0706.4469 v2.

\bibitem{chakraborty1} F.D.M. Haldane, Phys. Rev. Lett. 55, 2095 (1985); T. Chakraborty and P. Pietilainen, "The Quantum Hall Effects, Fractional and Integral" , second edition, Springer (1995) p163-172. 

\bibitem{chakraborty2} D. Yoshioka, B. I. Halperin, P. A. Lee, Phys. Rev. Lett. 50, 1219 (1983); T. Chakraborty and P. Pietilainen, "The Quantum Hall Effects, Fractional and Integral" , second edition, Springer (1995) p39-45.

\bibitem{white} S. R. White, Phys. Rev. B 48, 10345 (1993).

\bibitem{halperin} B. I. Halperin, Helv. Phys. Acta 56, 75 (1983).

\bibitem{yoshioka} D. Yoshioka, A. H. MacDonald, and S. M. Girvin, Phys. Rev. B 39, 1932 (1989).

\end{references}
\end{document}